# Self-guiding of 100 TW Femtosecond Laser Pulses in Centimeter-scale Underdense Plasma

L. M. Chen, H. Kotaki, K. Nakajima[†], J. Koga, S. V. Bulanov[‡], T. Tajima

*Kansai Photon Science Institute, Japan Atomic Energy Agency, Kyoto 619-0215, Japan*

Y. Q. Gu, H. S. Peng, X. X. Wang, T. S. Wen, H. J. Liu, C. Y. Jiao, C. G. Zhang,
X. J. Huang, Y. Guo and K. N. Zhou

*Laser Fusion Research Center, China Academy of Engineering Physics,
Sichuan 621900, China*

J. F. Hua, W. M. An, C. X. Tang, Y. Z. Lin

*Accelerator Laboratory of Tsinghua University, Beijing 100080, China*

An experiment for studying laser self-guiding has been carried out for the high power ultrashort pulse laser interaction with an underdense plasma slab. Formation of an extremely long plasma channel and its bending are observed when the laser pulse power is much higher than the critical power for relativistic self-focusing. The long self-guiding channel formation is accompanied by electron acceleration with a low transverse emittance and high electric current. Particle-in-cell simulations show that laser bending occurs when the accelerated electrons overtake the laser pulse and modify the refractive index in the region in front of the laser pulse.

---

[†] Also at High Energy Accelerator Research Organization (KEK), Ibaraki 305-0801, Japan
[‡] Also at A. M. Prokhorov Institute of General Physics RAS, Moscow 119991, Russia

Propagation of ultra-high intensity laser beams in plasmas has recently received much attention in connection with their potential applications for the development of x-ray lasers [1], plasma-based accelerators [2-4], and "fast-ignition" of precompressed thermonuclear fusion targets [5]. For all of these applications, it is necessary for the high intensity laser beam to propagate controllably over a long distance with high directionality. If the laser peak power is high enough, a laser beam can overcome the natural limit of refraction, and become focus in the plasma due to non-linear self-interaction [6, 7]. The balance between the self-focusing and diffraction can provide a condition for the long-distance propagation of the beam with peak intensity higher than otherwise achievable by focusing in vacuum. The self-guiding of a long enough single laser pulse in a plasma is known to be unstable [8]. The laser pulse can change directions of its propagation through hosing and bending. This problem has been addressed in theoretical papers [8], but so far no experimental evidence of laser pulse bending has been found.

In this letter, we present the first results of the experimental investigation of relativistically self-guided high power ultra-short pulse laser evolution in underdense plasmas in which the above mentioned bending regime is distinctly observed. Despite bending the laser radiation propagates over a very long distance in an underdense plasma and is accompanied by ultrarelativistic electron generation.

The experiment is carried out in the Laser Fusion Research Center with the SILEX-I laser [9], which is a 100 terawatt Ti:Sapphire laser working at a centre wavelength of 800 nm with a laser contrast of $10^7$. The pulse with a duration equal to $\tau_0$=30 fs was focused with a f /3 off-axis parabola onto a focal spot size $w_0$=16 μm. In the focal region the laser intensity is I=$6\times10^{18}$ W/cm$^2$ on average and $1.2\times10^{19}$ W/cm$^2$ at the peak, considering ≈50% of energy concentration in the focal spot. A supersonic pulsed slab He gas jet is used in order to form a spatially well defined gas target. The gas jet is produced by a slit nozzle, which is 1.2 mm wide and 10 mm long with a rectangular exit. After the nozzle, a beam charge transformer (BCT) is introduced in order to measure the electric charge of the accelerated electrons. A DRZ phosphor screen [10] is used to detect the transverse spatial profile of the electron beam. A bluepass filter (410 nm - 532 nm) is attached to a 16-bit CCD located in the laser polarization direction. It is used to measure the time-integrated plasma recombination fluorescence. A bandpass filter, with a bandwidth equal to 10 nm centered at 790 nm, is placed in the front of a CCD which is orientated perpendicular to the laser polarization and to the laser axis directions in order to measure the Thomson scattering of the laser light.

Figs. 1(a) to 1(c) represent the images obtained in the Thomson scattering measurements. The images are taken for different values of the plasma density when the laser pulse power is about 100 TW. We see in Figure 1(a), 1(b) that in the case of high gas pressure (p> 2.0 MPa) an extremely long plasma channel with a length approximately equal to ≈10 mm is formed. To our knowledge, this is the *longest* self-guided plasma channel observed till now stimulated by a single relativistic intensity laser pulse propagating in underdense plasmas. We note that in the moderate laser intensity range multi-kilometer distance pulse propagation in air, when the self-focusing due to the Kerr non-linearity is balanced by the plasma effects, has been observed in a number of the experiments [11]. In Figure 1(b), the laser diffraction cone is drawn to highlight the formation of this long plasma channel whose length is a factor > 20 larger than the Rayleigh length $Z_R= \pi w_0^2 /\lambda$, where $w_0$ is the laser spot size and $\lambda$ its wavelength. Under the experimental conditions $Z_R$ is about ≈ 400 μm. The Thomson imaging for higher plasma density, shown in Fig. 1(b), demonstrates a narrower column than for lower plasma density, presented in Fig. 1(a), along with weaker brightness. These correspond to the fact that the laser pulse with relatively higher energy is more effectively guided inside the plasma. For relatively low pulse energy and plasma density, as shown in Fig.1(c), only a short and wide scattering region is seen with a length comparable to the Rayleigh length, giving evidence that a major part of laser photons is scattered out of that region.

The change of the initial gas density and laser pulse energy has an expected effect on the self-focusing. In the case of an ideal Gaussian beam profile the critical power for self-focusing [12] is *$P_{cr}=16.2(n_{cr}/n_e)$GW*, where $n_e$ is the initial plasma density and $n_{cr}=m_e\omega^2/4\pi e^2$ is the critical density. When the backing pressure is above 2.0 MPa, the average plasma density exceeds $5\times10^{18}$ cm$^{-3}$ according to the calibration curve in our experiment nozzle. The experiment's 100 TW laser power is well above the self-focusing threshold. In Fig. 1(a) we see that the emitting plasma is confined inside a narrow column, exhibiting pulse self-focusing. For lower plasma density and lower pulse energy pronounced self-focusing is not observed. This can be explained as follows: for short enough pulses, much shorter than the wake wave wavelength, the self-focusing does not develop because the summarized plasma collective response is not strong enough to substantially modify the refractive index. In this regime the front part of the pulse expands over the distance of the order of the diffraction length $Z_R$. In Ref. [13] it is shown that when the laser pulse is too short, self-focusing does not occur, even for powers greater than $P_{cr}$ (see also computer simulation results presented in Ref. [14]). A roughly estimated threshold for the laser pulse length needed to self-focus can be written as $l_{las} \geq \lambda_{wf}$, where $\lambda_{wf} \approx 2\pi c/\omega_{pe}$ is the wake wave wavelength. For the average plasma density $5\times10^{18}$ cm$^{-3}$, the laser pulse length, $l_{las}$=12.5 μm, is of the order of the Langmuir wave wavelength, ($\lambda_{wf} \approx 13.5$ μm), i.e. the laser pulse undergoes self-focusing over the length $\approx Z_R(P_{cr}/P)^{1/2} \approx 90$ μm. Our experiment demonstrates that the laser radiation is subject to self-focusing when the condition $l_{las} \geq \lambda_{wf}$ is fulfilled. On the other hand, if $l_{las} << \lambda_{wf}$ although the laser pulse power in the focal spot, 100 TW, is well above the critical power for self-focusing when the backing gas pressure is continuously reduced to 1.5 MPa, the averaged plasma density quickly decreases to $\approx 2\times10^{18}$ cm$^{-3}$ and it results in $l_{las}/\lambda_{wf}$=0.6. Under this condition laser pulse self-focusing becomes much more difficult due to defocusing caused by the electron density maximum at the front part of the wake balancing the relativistic focusing, as pointed out in Refs. [13, 14]. This might be a reason for the observed plasma miss-channeling in the case of low plasma densities.

In addition to the observation of the long plasma channel, the propagation of the laser pulse in fact becomes dominated by the long scale bending as revealed in PIC simulations presented in Refs. [15]. The self-focusing channel bending has been observed in our experiments by employing the Thomson scattering measurements (see Fig. 1(a), 1(b)), and by using plasma fluorescence imaging (this is shown in Fig. 2). We see that the channel is bent significantly to the right side away from laser incident direction (top-view in Figure 1) in the plane perpendicular to the laser pulse polarization, and guided randomly shot by shot in the plane of the laser pulse polarization, as seen in Fig. 2. We mention that the laser pulse does not show bending with the laser pulse incident onto plasmas in the low energy limit when we check the CCD in the laser propagation direction. This excludes the possibility of laser misalignment or pointing vibration.

In order to unravel the physical mechanism for the self-focusing channel bending, we carry out 2D particle-in-cell simulations of the interaction of a laser pulse with underdense plasma with the use of the REMP code [16]. In the cases presented here linearly p- and s-polarized laser pulses, with the electric field in the y- and z-directions, respectively, propagate along the x-axis. The laser pulses with a maximum dimensionless amplitude $a_0$=3 have a Gaussian envelope with FWHM size 10×16$\lambda^2$. The plasma density is $n_0$=2×10$^{-2}n_{cr}$ which corresponds to the ratio $\omega_{pe}/\omega$=0.1412. The ions have an absolute charge equal to that of the electrons, and mass ratio $m_i/m_e$=1836. The simulation box has 10050×450 grid points with a 0.125$\lambda$ mesh size. The target has the form of an underdense plasma slab of size 1250×50$\lambda^2$. The total number of quasiparticles is about *10$^8$*. The boundary conditions are absorbing in all directions for both the electromagnetic (EM) radiation and the quasiparticles. The space and time units are the wavelength $\lambda$ and the period *2π/ω* of the incident radiation. In these simulations we have chosen a plasma density more than an order of magnitude higher than in the experiment in order to have conditions which provide the laser pulse self-focusing and bending over the distance no greater than a millimeter, because full-scale detailed simulations of a one *cm* scale plasma length takes time and computer resources.

The simulation results are presented in Figs. 3 and 4. In both cases of the p- and s-polarized pulses the simulations show formation of a long and narrow self-focusing channel with its length over hundreds of $\lambda$. The laser pulse undergoes self-focusing, as seen in Figs. 3 (a) and 4 (a). In Fig. 3 (a) we present the the z-component of the magnetic field of the p-polarized pulse in the x,y plane at t=250 2π/ω. Fig. 4 (a) shows the z-component of the electric field of the s-polarized pulse in the x,y plane at t=250 2π/ω. The pulse generates a wake field which accelerates the electrons injected due to transverse wake wave breaking [17] (see Figs. 3 (b) and 4 (b), where the electron density distribution in the x,y plane at t=250 2π/ω is shown). In the case of the p-polarized pulse the wake is asymmetric at the front, Fig. 3 (a), and shows hosing, i. e. there are low frequency oscillations in the transverse direction with the wavelength along the pulse propagation direction equal to $\lambda c/(v_{ph}-v_g)$ with $v_{ph}$ and $v_g$ being the phase and group velocities of the pulse, which for the simulation parameters gives 30 $\lambda$ (see theoretical description of the hosing in Ref. [15]). Over a long time scale, around t=400 2π/ω, energetic electrons, accelerated by the wake field in the *x*-direction overtake the pulse and due to the relativistic dependence of the electron mass on the energy, modify the refractive index in the front of the pulse [15,18]. In turn the effect of the initial deflection of fast electrons behind the pulse front is transferred by the electrons themselves to the plasma in front of the pulse. This results in an asymmetric change of the refractive index in front of the pulse, which causes the deflection of the pulse and reinforces the asymmetric acceleration of fast electrons inside the pulse. Simulation results of the laser pulse with both the *p*- and *s*-polarizations demonstrate that the self-focusing is asymmetric with some filaments in the leading part of the laser pulse (see Figs. 3 (d) and 4 (d)).

Our observations show that plasma emission emanates from a number of bright hollow spots situated periodically along the center axis, or along the laser bending directions in the vertical plane. It shows an "atoll" structure and there is a density valley in its center, as seen in Fig. 2. This phenomenon can be explained by paying attention to the formation of relativistic electron vortices, associated with a quasistatic magnetic field, which are clearly seen in Fig. 3 (a), and a relativistic soliton train behind the laser pulse (see Fig. 4 (a)). Relativistic electron vortices and electromagnetic solitons are basic entities generated in the nonlinear laser plasma interaction (e. g. see review article [20]). Inside the soliton a low frequency high amplitude electromagnetic mode is trapped. The ponderomotive pressure of this electromagnetic mode displaces electrons outward and the Coulomb repulsion in the electrically nonneutral ion core pushes ions away. As a result, bubbles in the ion density distribution are formed [21]. In the case of an electron vortex the cavity is formed by the quasistatic magnetic field pressure. In Figs. 3 (c) and 4 (c), where the ion density distribution is shown, we clearly see the formation of cavities inside the self-focusing channel.

Accelerated electron charge and emittance have also been studied experimentally in the case of the long plasma channel formation. The measured electric charge of the accelerated electrons ($E >1$ MeV) is found to be 10 nC per shot, which is the highest value for laser accelerated electrons to the best of our knowledge. This electron bunch is tightly collimated with an emittance $\approx 0.8\pi$ mm mrad. However, for the shorter plasma channel formation, the maximum accelerated electron charge is dramatically reduced to 0.5 nC, in conjunction with generation of a less collimated beam with emittance $\approx 4\pi$ mm mrad, see Fig. 5. When the long plasma channel is formed, we see quasi-monoenergetic electron bunch generation at 70 MeV with a well defined energy spread of 10 MeV in FWHM. In addition, in the regimes of strong laser pulse bending the fast electron bunch is also bent in the same direction as the laser pulse as seen in Fig. 5.

In summary, first experiments aimed at studying laser self-guiding in a long slab underdense plasma have been performed with 30 fs, 100 TW laser pulses. We observe a 10 mm length plasma channel formation. In the case of laser pulse guiding inside the long channel a quasi-monoenergetic electron bunch is generated with very low emmitance and high electric charge. The laser bending is ascribed to accelerated electrons overtaking the laser pulse and changing the refractive index in front of the pulse.

This work was jointly supported by JST trilateral cooperation funding and Kakenhi project in KPSI, JAEA.

Figure Captions

Fig. 1 (color): Top-view of the Thomson scattering measurement. (a) the laser energy is 2.5 J, the gas pressure 2 MPa; (b) the laser energy is 3.0 J and the gas pressure equals 2.5 MPa; (c) the laser energy is 1.9 J, the gas pressure equals 1.5 MPa.

Fig. 2 (color): The side-view plasma imaging in the case of the laser energy equal to 2.5 J and the gas pressure equal to 2 MPa.

Fig. 3 (color): Distribution of the z-component of the magnetic field, with the inserted magnetic field image behind the pulse plotted with higher resolution, (a); electron density (b) of the p-polarized pulse in the x,y plane at t=250 2π/ω; the ion density at t=500 2π/ω (c), and the ion distribution (stretched in the vertical direction) at t=1112.5 2π/ω (d).

Fig. 4 (color): Distribution of the z-component of the electric field, with the inserted electric field image behind the pulse plotted with higher resolution, (a); electron density (b) of the s-polarized pulse in the x,y plane at t=250 2π/ω; the ion density at t=500 2π/ω (c), and the ion distribution (stretched in the vertical direction) at t=1250 2π/ω (d).

Fig. 5 (color): (left) Transverse spatial distribution of fast electrons with the energy above *1 MeV* in case of long plasma channel formation. The short *(≈3 mm)* guided accelerator produced weakly collimated electron beams (middle). Imaging plate (right), placed close and facing the laser propagation direction, shows evidence of electron bending along laser pulse (white circle shows laser pointing direction).

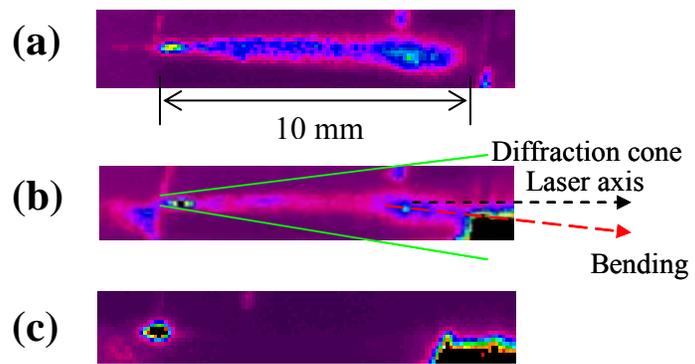

Fig. 1 (color): L. M. Chen et al.

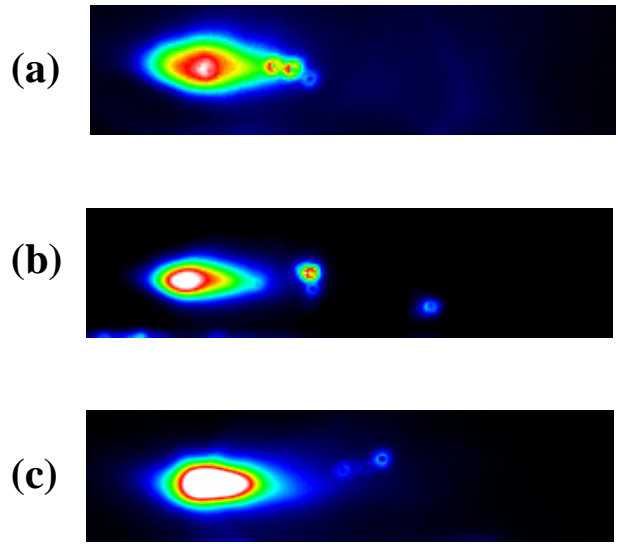

Fig. 2 (color): L. M. Chen et al.

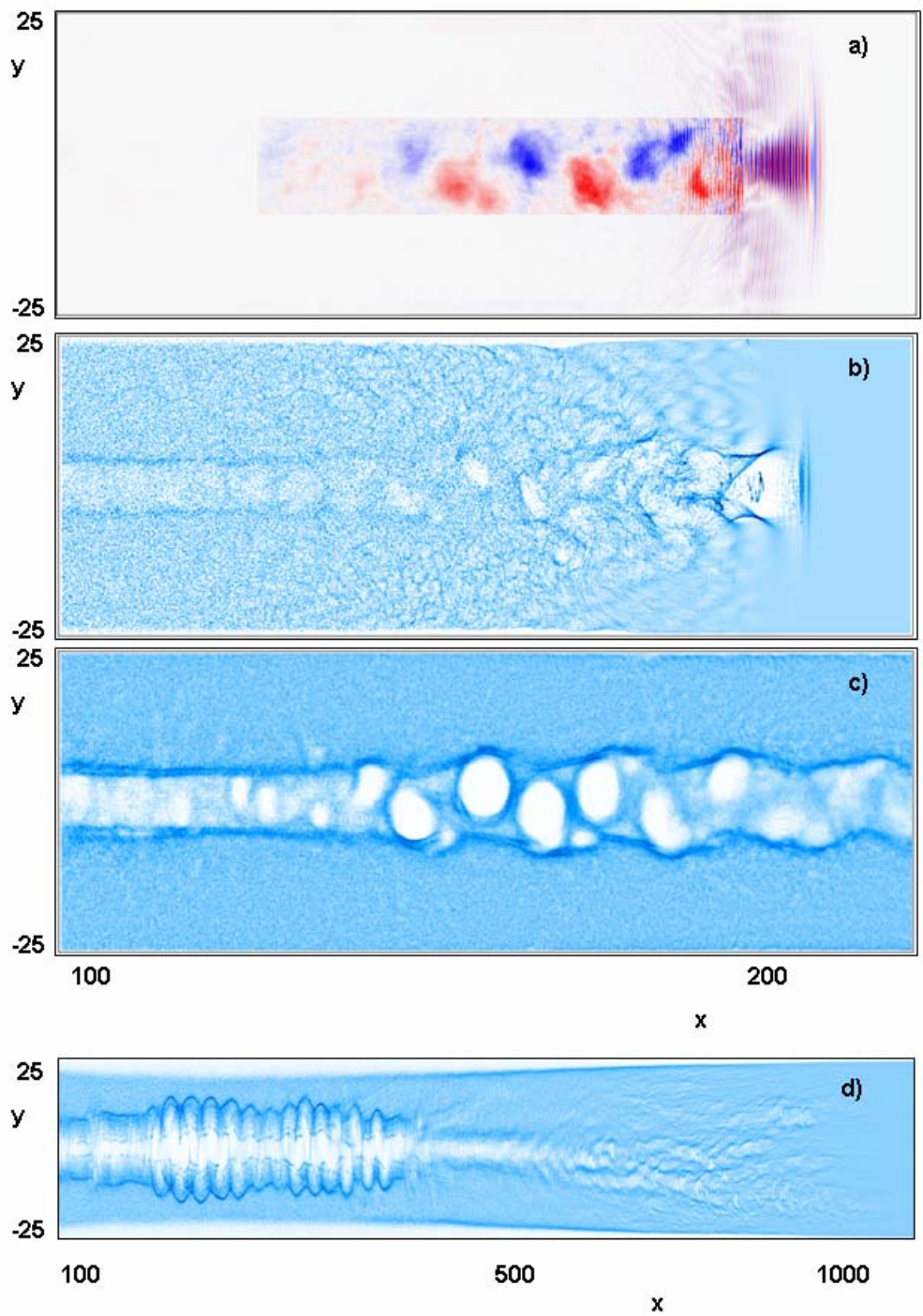

Fig. 3. (color): L. M. Chen et al

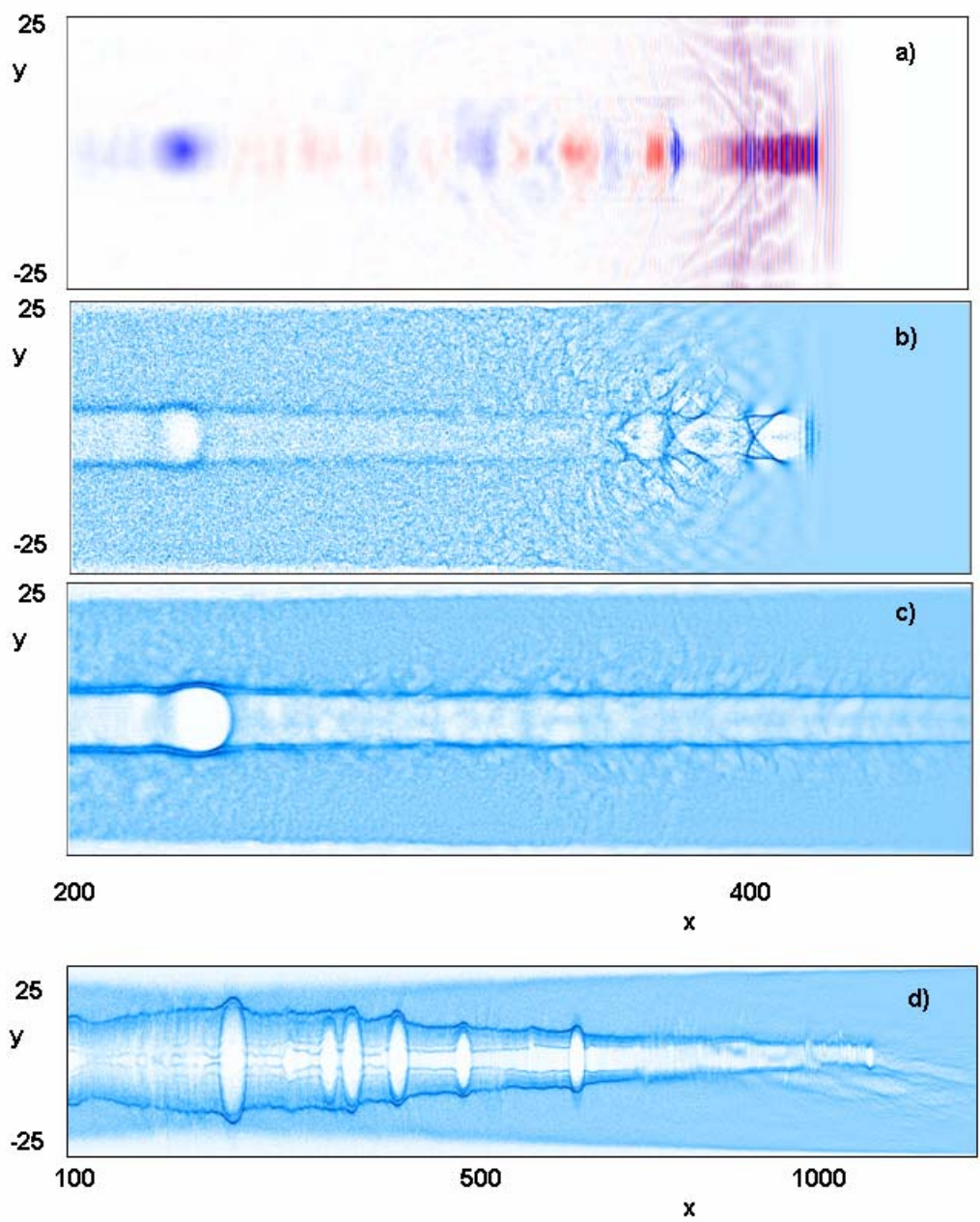

Fig. 4. (color): L. M. Chen et al

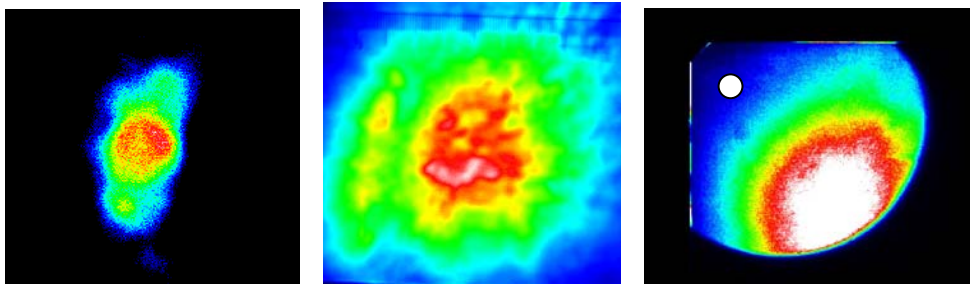

Fig. 5 (color): L. M. Chen et al.